# Integration of deep learning with expectation maximization for spatial cue based speech separation in reverberant conditions

Sania Gul[1], Muhammad Salman Khan[2*], Syed Waqar Shah[1]

[1]Department of Electrical Engineering, University of Engineering and Technology, Peshawar, Pakistan.

[2]Department of Electrical Engineering, (JC), University of Engineering and Technology, Peshawar, Pakistan.

**Abstract:** *In this paper, we formulate a blind source separation (BSS) framework, which allows integrating U-Net based deep learning source separation network with probabilistic spatial machine learning expectation maximization (EM) algorithm for separating speech in reverberant conditions. Our proposed model uses a pre-trained deep learning convolutional neural network, U-Net, for clustering the interaural level difference (ILD) cues and machine learning expectation maximization (EM) algorithm for clustering the interaural phase difference (IPD) cues. The integrated model exploits the complementary strengths of the two approaches to BSS: the strong modeling power of supervised neural networks and the ease of unsupervised machine learning algorithms, whose few parameters can be estimated on as little as a single segment of an audio mixture. The results show an average improvement of 4.3 dB in signal to distortion ratio (SDR) and 4.3% in short time speech intelligibility (STOI) over the EM based source separation algorithm MESSL-GS (model-based expectation-maximization source separation and localization with garbage source) and 4.5 dB in SDR and 8% in STOI over deep learning convolutional neural network (U-Net) based speech separation algorithm SONET under the reverberant conditions ranging from anechoic to those mostly encountered in the real world.*

**Keywords**: machine learning; deep learning; reverberations; interaural cues; perceptual techniques.

## I. Introduction

Blind source separation (BSS) is the process of separating a signal of interest from a mixture. There are many day to day situations where a signal needs to be separated from the mixture e.g. the satellite images and the radar signals need to be recovered from the clutter for accurate target positioning, the electroencephalogram (EEG) needs to be cleared from the noise generated by the muscular activity in brain, and the communication signal needs to be recovered from the background noise and interferences.

--------------

*Corresponding author: Muhammad Salman Khan, University of Engineering and Technology (JC), Peshawar, Pakistan. E-mail address:* salmankhan@uetpeshawar.edu.pk

Audio source separation is one of the broad fields of blind source separation which covers only the audio signals. This field is further divided in two vast fields of speech separation and music separation.

While human beings with normal hearing are capable of separating the signal of interest in noisy and reverberant environment without any trouble, the machines and the people suffering from hearing impairment are incapable of doing so.

With many applications of machine listening emerging rapidly, efficient audio source separation is the need of time. These applications cover almost every sphere of modern living, spanning, for instance, from the home to the office, from security to health care and even in the fun-filled activities. Audio source separation is useful, for instance, in communication for automatic speech recognition (ASR), internet voice search engines and home assistants, in medical engineering for designing better hearing aids, in music for transcription of musical records, in offices for implementing smart meeting rooms which record minutes of meeting automatically, and for safety and security purposes to detect anomaly from sounds.

In the past, many machine learning algorithms e.g. maximum likelihood estimation (MLE) [1], expectation maximization (EM) [2], and support vector machine (SVM) [3] had been used for speech separation task. Although they do not require lengthy trainings, as required by deep learning, they are restricted in terms of performance. Introduction of deep learning in speech separation has made possible tremendous improvement in the separation performance which was earlier unimaginable with the machine learning algorithms.

Recent years have witnessed the increase in use of deep neural networks either alone e.g. [4], [5] and [6] or in combination with the machine learning algorithms e.g. [7], [8], [9] and [10]. In [7], unsupervised spatial clustering is done by complex angular central Gaussian mixture model (cACGMM) which is then used to guide the training of a deep clustering single channel spectral network. In [8], the same technique of [7] is used but instead of cACGMM, the von-Mises-Fisher complex angular central Gaussian mixture model (vMF-cACGMM) and the Gaussian complex angular central Gaussian mixture model (G-cACGMM) are used where each of them is integrated individually with both the deep clustering (DC) and the deep attractor network (DAN). In both [7] and [8], the machine learning algorithms are used for clustering multi-channel spatial cues while the neural network is used for clustering single channel spectral cues. In [9] also, deep neural network (DNN) is used to model the spectral information and the classical multichannel Gaussian model is used to exploit the spatial information. The framework is built upon the state-of-the-art iterative EM algorithm which integrates these spatial and spectral models in a probabilistic fashion.

Along with speech, music source separation (MSS) has been an active research area over the last decade. The goal is to design algorithms that can separate vocals and instruments. These algorithms have been successfully used for rendering music, instrument generation for Karaoke systems, or singer identification. Unsupervised machine learning methods, have difficulty in capturing the dynamics of vocals and instruments, while supervised methods, suffer from the generalization and processing speed issues. In [10], conventional regression based mask inference clustering (a machine learning algorithm) is integrated with deep learning spectral network to achieve better separation of vocals from the instruments. Whereas conventional mask-inference approach only focuses on increasing the separation between the sources, the deep clustering reduces the within-source variance. It is shown in [10] that the new hybrid network outperformed both the plain deep clustering network and the conventional mask inference network.

In all of the methods discussed above, the integration of machine learning with deep learning has significantly outperforms either of its components, motivating the fusion of the best techniques of both paradigms to complement each other in the final goal of better source separation. The only thing common for all the integration techniques discussed above is that they are using neural networks for spectral cues and machine learning for spatial cue clustering.

Most of the DNN based BSS algorithms lack the capability to utilize the multichannel acoustic cues [9]. Most of the DNN approaches are single channel source separation techniques, where the input signal is either one of the channels of the original multichannel mixture signal or the output of delay-and-sum (DS) beamforming. In order to take the benefit of multichannel cues, they are dependent on machine learning algorithms.

Our proposed methodology of integration is different from those discussed above, as it integrates machine learning with deep learning for only the spatial cue information. To the best of our knowledge, this is the first time such merger of the two standards is proposed for the spatial cue based source separation systems.

The rest of the paper is organized as follows. In the next section, we will focus on the related work. We will give our proposed system overview in section III. In section IV, we will present experimental details including the room layout and the dataset used. The experimental results and comparison of different algorithms is given in section V and the paper is concluded in section VI.

## II. Related Work

Speech separation by interaural cues (ILD and IPD) has achieved tremendous success due to its similarity with the psychoacoustic approach of speech separation [1] and [2]. Both ILD and IPD cues are utilized simultaneously for better separation. This is because of the fact that the IPD cues are ambiguous at higher frequencies when the microphone spacing is larger than the half wavelength due to the phase wrap problem [1]. Under such conditions, the ILD becomes indicative about target directions [9]. However, as the gains at different microphones are usually very similar under the far field assumption [5], the ILD is considered to be a weak cue for speech separation tasks, especially below 3 to 4 kHz. So, these interaural cues are used simultaneously to complement each other across the entire speech band.

MESSL [2] is an interaural cue based speech separation model based on windowed disjoint orthogonality (WDO) assumption [12], which states that no two speech sources occupy a single TF point simultaneously [1]. MESSL uses expectation maximization (EM) algorithm (a machine learning algorithm) for speech separation. The EM algorithm is a two-step iterative process, in which the process switches back and forth between Expectation (E) and Maximization (M) steps until the, "best fit" (Maximum Likelihood (ML)) parameters are generated for each source. The convergence point is reached either if the model parameters stop changing from one iteration to the next or the program runs up to the maximum iteration setting. EM algorithm is sensitive to initialization parameters of each cluster and requires the knowledge about the number of clusters (i.e. the number of sources) in the mixture, and the distribution type (Gaussian, uniform, exponential etc.) of each cluster.

MESSL assumes Gaussian distribution for the each source in the audio mixture. As already discussed, due to the sensitivity of EM algorithm to initialization, MESSL uses highly accurate PHAT algorithm [13] for the initialization of its IPD model. The use of both ILD and IPD cues, PHAT algorithm for IPD initialization and removal of phase wrap problem by using the top down approach of estimating the IPD from the interaural time difference (ITD) instead of directly observing them from the audio mixtures, results in highly improved speech separation performance which was not possible when EM was used for clustering only the ILD cues [14].

In order to deal with reverberations, MESSL uses an additional functionality called the garbage source (GS). GS is a virtual source, having a uniform distribution. Any outlier spatial cue, far off from the mean values of ILD and IPD of the real sources is allocated to GS, resulting in their cleaner outputs, which would otherwise be contaminated by the presence of outliers. The version of MESSL with GS activated is called MESSL-GS.

MESSL estimates soft (probabilistic) ILD and IPD mask for each source separately by using the EM algorithm and applies their product on the audio mixture's spectrogram to retrieve the sources.

SONET [11] is a U-Net (deep learning convolutional neural network) based interaural speech separation model, designed for anechoic conditions. Like MESSL, this model is also based on WDO assumption and uses the spatial cues for source separation. In SONET, two separate U-Nets (a specialized neural network for semantic segmentation) are trained on the interaural level difference (ILD) and the interaural phase difference (IPD)) spectrograms generated by a single source. These U-Nets are named as SONET-L (SONET trained on ILD cues) and SONET-P (SONET trained on IPD cues) in [11]. After training, these U-Nets are used to predict the class of each time frequency (TF) unit of the interaural spectrogram of an audio mixture. The output of SONET-L is treated as the soft ILD mask and the output of SONET-P is treated as the soft IPD mask for each source in the audio mixture. Instead of taking the product of ILD and IPD masks as employed by MESSL, SONET concatenates different portions to ILD and IPD masks based on their strength in different speech bands. These portions are further weighed empirically for better source retrieval. Such soft mask was termed as 'weighted sub-band mask' in [11].

Figure 1 highlights some basic differences in the operation and soft masks of MESSL and SONET.

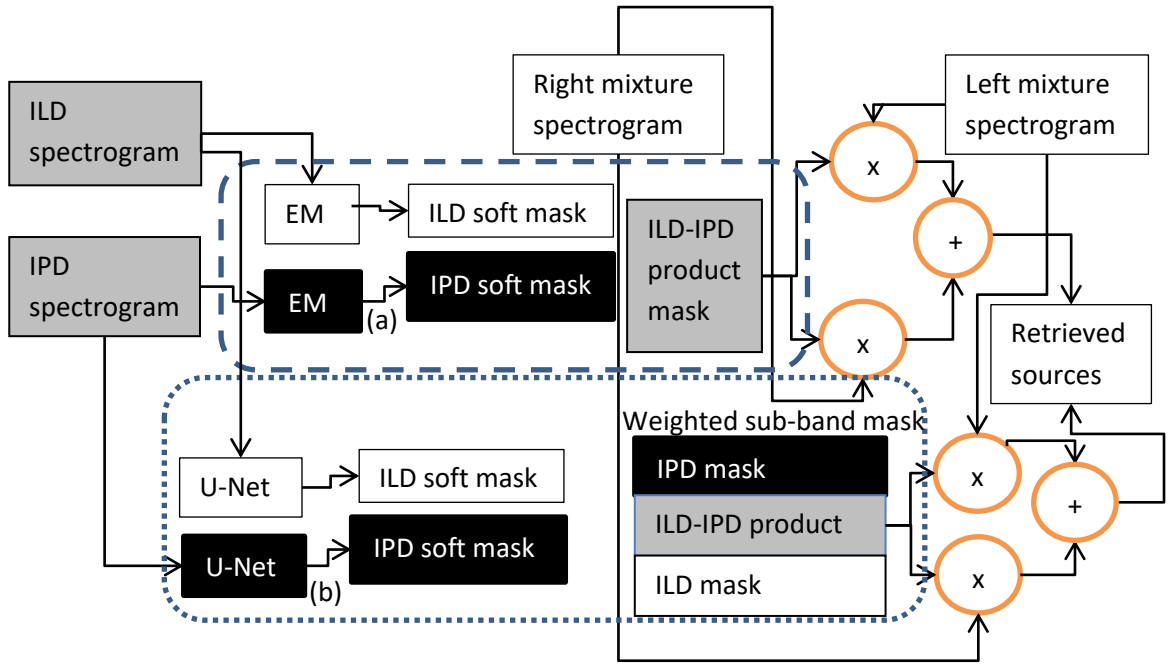

**Figure 1:** Highlighting the differences between the operation of SONET and MESSL. (a) shows MESSL and (b) shows SONET. The encircled 'x' shows the element wise product and the encircled '+' element wise summation operation. The weights of 'Weighted sub-band mask' are as defined in Table 3.

Also, as the training phase of deep neural network is lengthy and requires high-end computational resources (graphical processing unit (GPU)), a 'generalized solution' is suggested in [11] as shown in Figure 4 (b), where the network trained for one source position can be used for other sources placed in its neighboring positions due to similarity of interaural cues generated by these sources.

SONET performs very well when using only the ILD cues, but it cannot discriminate the TF units on the basis of their IPD cues very accurately. This is due to its inability to tackle the phase wrap problem. Removing the phase wrap problem by using the top down approach of [2] did not work well in case of U-Net, as it reduces the IPD variance of each source. This variance reduction works well with the expectation maximization (EM) algorithm [2] but not for the convolutional neural network U-Net [11]. The inclusion of IPD cues (whether the values observed from the mixture or those after the phase unwrap by the top down approach) in SONET [11], resulted in decline of its output performance. The performance comparison of the two speech separation models, one using the EM algorithm and the other using the SONET-P network for clustering the IPD cues, is given in the 'experiment' section (Section V).

Since the strengths of deep clustering and conventional machine learning architectures appear complementary for ILD and IPD cues, we explore combining them in a single hybrid network which uses the U-Net based ILD cue discrimination with the EM based IPD cue discrimination to achieve more efficient speech separation. In the next section, we explain the working of our proposed algorithm.

## III. Proposed Methodology

The block diagram of our proposed system is shown in Figure 2. We have named our proposed system 'EM-SONET (EM-S)', depicting the fusion of EM with U-Net based pre-trained speech separation neural network (SONET) for clustering the interaural cues.

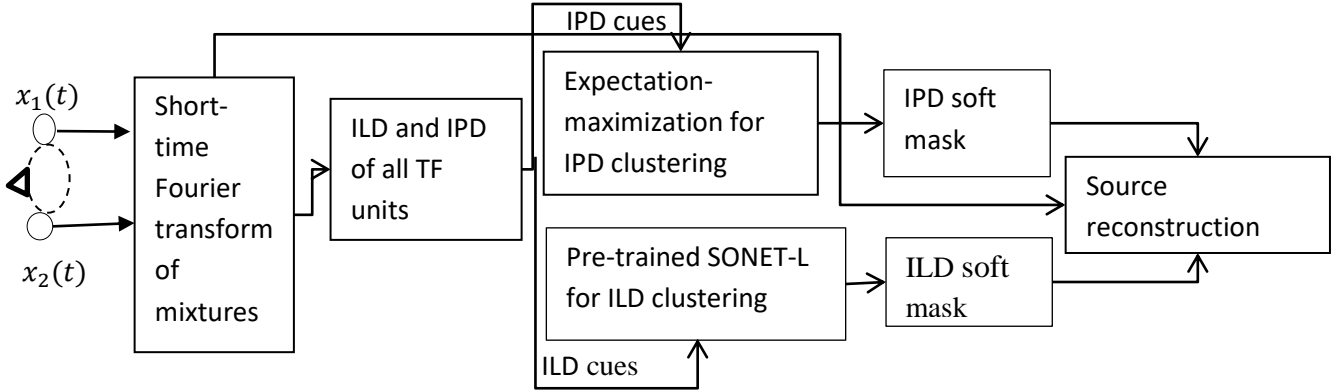

**Figure 2: Block diagram of EM-SONET (EM-S)**

Assume that there are $Q$ active sources present inside the room and there are $k$ microphones, recording the audio mixtures (where k = {1, 2} for binaural speech separation). Each source $s_i$ is convolved with the room impulse response (RIR) $h_{ki}$ that exists between the source $s_i$ and the $k^{th}$ microphone and added together with other convolved sources to form mixture $x_k$ at the $k^{th}$ microphone. The process of mixture formation at the $k^{th}$ microphone is given by equation (1) and the signal notations are shown in Figure 3:

$$x_k(t) = \sum_{i=1}^{Q} s_i(t) * h_{ki}(t) \qquad (1)$$

where '*' represents the convolution operation and $t$ represents the sampling time index when the sampling frequency at which mixture samples are taken is $f_s$. The assumption is made that except for the $Q$ active sources, there are no unidentified noise generators which are contributing to the mixtures.

Short-time Fourier transform (STFT) of the signal $x_k$ collected at the $k^{th}$ microphone is taken after segmenting $x_k$ into overlapping frames of length $L$ and windowing each frame with the window function $w(t)$ as given in equation (2).

$$X_k(\omega, m) = \mathcal{F}(w(t)x_k(t)) \qquad (2)$$

Where $\mathcal{F}$ is the STFT operator and $w(t)$ is the hamming window function given as $w(t) = 0.54 - 0.46\cos(2\pi t/N), 0 \leq t \leq N, where\ N = L - 1$, and $L$ is the window length (See Table 2 for STFT parameters). $\omega$ is the discrete circular frequency index and $m$ time frame index in the STFT domain.

The ratio of the STFT of signals at both microphones is taken to generate the interaural spectrogram as given in equation (3).

$$\frac{X_1(\omega,m)}{X_2(\omega,m)} = \alpha(\omega,m)e^{i\phi(\omega,m)} \qquad (3)$$

Where $\alpha(\omega,m)$ is the ILD and $\phi(\omega,m)$ is the observed IPD at a TF point having discrete circular frequency index $\omega$ and time frame index $m$.

Converting the ILD to decibels (dB) at each TF point by the formula in equation (4), we obtain the ILD spectrogram of the audio mixture.

$$ILD(dB) = 20log_{10}\alpha(\omega,m) \qquad (4)$$

This ILD spectrogram is given at the input of a pre-trained model SONET-L which produces the ILD soft mask for each source at its softmax layer. SONET [11] is a deep learning based algorithm using semantic segmentation of the interaural spectrograms of audio mixtures for speech separation.

The IPD cues of the mixtures are given to the expectation maximization block. The observed IPD values from the mixtures i.e. $\angle \frac{X_1(\omega,m)}{X_2(\omega,m)}$ at each TF point (obtained from equation (3)) do not always map to the correct interaural time difference (ITD) due to spatial aliasing. So a top down approach inspired by [2] is used for the calculation of IPD, where IPD is estimated by plugging in different values of ITD (τ) in the range from -15 to 15 samples in 0.5 sample increments. At the sampling frequency of 16 kHz, this range corresponds to around ± 1 milliseconds in 31.25 micro-seconds increments. The $\tau$ which produces the closest match to the observed IPD is selected. However, it is required that the delay ($\tau$) and the length of RIR must be smaller than the STFT frame length. Any portion of RIR above the STFT frame length would be treated as noise.

The phase residual error $\hat{\phi}$ is defined as the difference between the observed IPD and the estimated IPD and given in equation (5) as:

$$\hat{\phi} = \angle \frac{X_1(\omega,m)e^{-j\omega\tau}}{X_2(\omega,m)} \qquad (5)$$

$\hat{\phi}$ lies in the interval {-π, π}. The IPD residual is modeled as normal distribution. Let $\xi(\omega)$ and $\sigma^2(\omega)$ be the mean and variance of the IPD residual ($\hat{\phi}$). Then the IPD model for each source $s_i$ at each TF point is given in (6) as:

$$p(\hat{\phi}(\omega,m;\tau)|\hat{\theta}) = \mathcal{N}\left(\hat{\phi}(\omega,m;\tau)\big|\xi_{i,\tau}(\omega),\sigma_{i,\tau}^2(\omega)\right) \qquad (6)$$

The subscripts with mean and variance symbols in equation (6) show that the IPD parameters are dependent on both frequency $\omega$ and delay $\tau$. And

$$\hat{\theta} = \{\xi_{i,\tau}(\omega),\sigma_{i,\tau}^2(\omega),\psi_{i,\tau}\} \qquad (7)$$

represents all model parameters for source $s_i$. $\psi_{i,\tau}$ is the mixing weight; i.e. the proportion of the total TF points of mixture belonging to source $s_i$ at delay $\tau$ in the Gaussian Mixture model (GMM), which manifests itself due to mixing of many such Gaussian distributions belonging to different combinations of sources and delays. The log likelihood $L$, given the observation $\hat{\boldsymbol{\theta}}$, over all TF points is given in (8) as:

$$L(\hat{\boldsymbol{\theta}}) = \sum_{\omega,m} \log p(\phi(\omega,m)|\hat{\boldsymbol{\theta}}) \quad (8)$$

Marginalizing over all sources and all delays, the log likelihood function is given as in (9).

$$L(\hat{\boldsymbol{\theta}}) = \sum_{\omega,m} \log \sum_{i,\tau} [\mathrm{N}\left(\hat{\phi}(\omega,m;\tau)\big|\xi_{i,\tau}(\omega), \sigma_{i,\tau}{}^2(\omega)\right) . \psi_{i,\tau}] \quad (9)$$

The maximum likelihood solution is given as in (10).

$$L(\hat{\boldsymbol{\theta}}) = max_\theta \sum_{\omega,m} \log p(\phi(\omega,m)|\hat{\boldsymbol{\theta}}) \quad (10)$$

This solution is achieved by using the expectation maximization (EM) algorithm, where the system switches back and forth between the E and the M step, until the maximum likelihood solution is obtained or the processor completes the pre-set number of iterations. In the E step, likelihood $v$ of each TF point belonging to source $i$ and delay $\tau$ is given as in (11)

$$v_{i,\tau}(\omega) \propto \psi_{i,\tau}.\mathrm{N}\left(\hat{\phi}(\omega,m;\tau)\big|\xi_{i,\tau}(\omega), \sigma_{i,\tau}{}^2(\omega)\right) \quad (11)$$

In the M step, new model parameters $\hat{\boldsymbol{\theta}}$ are estimated from all TF points, weighted by their corresponding likelihood probabilities (calculated in E step) by the formulae given in (19) to (23) in [2]. At the end, the probabilistic IPD mask for each source is given as in (12)

$$M_i(\omega,m) = \sum_\tau v_{i,\tau} \quad (2)$$

The ILD soft mask from SONET-L and the IPD soft masks from EM algorithm are combined to form sub-band mask as listed in Table 3 and given by equation (13).

$$Sub - band\ mask = \begin{bmatrix} IPD\ mask\ (0\ to\ 1.5\ kHz); \\ Product\ of\ ILD\ and\ IPD\ masks\ (1.5\ to\ 4\ kHz);\ ILD\ mask\ (4\ to\ 8\ kHz) \end{bmatrix} \quad (13)$$

This sub-band mask is formed on the basis of the strength of spatial cues in different portions of speech band. This mask is applied separately on each mixture's spectrogram $X_k$, and the results are added together and converted back to time domain signal by taking the inverse short-time Fourier transform (ISTFT).

Care must be taken when combining the IPD and the ILD masks by equation (13), as the sources are sometimes permuted by the EM algorithm [8]. For this purpose, we will use a technique mentioned in section IV j "Solving the EM permutation problem".

*Algorithm summary*

**Input:** Audio mixture.

**Output:** Target source.

1. Convert each audio mixture collected at the binaural microphone setup to time frequency (TF) domain by using equation (2).
2. Extract the IPD and the ILD spectrograms from these mixtures by using equations (3) - (5).
3. Apply the EM algorithm (equations (6) – (12)) on the IPD spectrogram to obtain the IPD soft mask for each source.
4. Apply the ILD spectrogram at the input of pre-trained SONET-L to obtain the ILD soft mask for each source.
5. Combine ILD and IPD soft masks of the target by using equation (13) after using the technique mentioned in section IV j.
6. Apply the sub-band mask obtained from equation (13) to each TF mixture individually and add the results.
7. Convert the sum back to time domain by ISTFT to retrieve the target.

## IV. Experimental Evaluation Parameters

### a. Room Layout

We perform experiments with two sources in each room; one is target and the other is masker. The masker is placed in front of microphones while the target is moving in the range $\{0^0 : 90^0\}$ with the increment of $15^0$, as shown in Figure 3. According to the room impulse responses (RIRs) used for our experiments, the distance between the two microphones is 175mm, which is equal to the average distance between the two ears of human beings and the radial distance between each source and the microphone setup is 1.5 m.

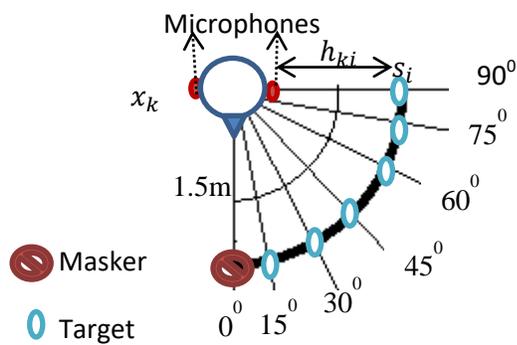

**Figure 3: Masker and all possible target positions inside the room**

The target and masker positions are kept according to the source positions of [11] as we are using the pre-trained neural network SONET-L of [11] for clustering the ILD cues, and this network was trained with the target and masker positions as shown in Figure 3. Other positions are not tried for target and masker, as the interaural cues vary with the source location and any new arrangement of them would require re-training of SONET-L which was not feasible to us due to computational constraints.

### b. Data set

We have tested five mixtures for each position in all the rooms mentioned in Table 1. Each mixture is made of two speech sources of 1.05 seconds duration taken from the TIMIT dataset [15]. Each source is normalized and convolved with the binaural RIR [16] according to its position in the room before being added together to form an audio mixture. The sampling frequency is 16 kHz.

As the number of active sources in each mixture is equal to the number of microphones, so this problem is the determined de-mixing blind audio source separation (BASS) problem. The reason of not exceeding the number of sources above two is that the pre-trained network SONET-L which will be used for clustering the ILD cues is trained for separating only two simultaneous active sources.

After convolution with the binaural room impulse response (BRIR), the duration of speech sources and the mixture resulted by the summation of such sources would increase beyond 1.05 seconds (depending on the reverberating time ($RT_{60}$) of the room). However, in order to keep the reverberations in the signals intact and evaluate the performance of the proposed system under reverberant conditions, we will not clip it off back to 1.05 seconds. This would not affect any of our processing steps afterwards, as the SONET-L can handle images (which results from conversion of time domain signal to time frequency (TF) domain) of dimensions larger than those over which it was trained. Neither is there any problem to the EM algorithm by the increase in signal duration due to reverberations.

### c. Binaural room impulse response (BRIRs)

We will use the RIRs of [16]. The $RT_{60}s$ of these rooms are in the range {0, 1} seconds, since these $RT_{60}s$ are representative of acoustic conditions mostly encountered in homes, offices, class rooms, studios, cinemas, lecture halls and conference rooms. These RIRs belong to the five rooms {X, A, B, C, D} and were captured by keeping the distance of 1.5 m between the loudspeaker and HATS (Head and Torso Simulator). They were obtained from the sine sweeps replayed through a Genelec 8020A active loudspeaker and the responses were deconvolved to produce the impulse responses. The recordings were made at a sampling frequency of 48 kHz (which was subsequently resampled to 16 kHz). The acoustic center of the loudspeaker was placed at the same height as the ears of HATS.

The anechoic RIRs of room X are generated by pseudo-anechoic method, where the responses captured in a large room were simply truncated before the first reflection, so that subsequent reflections did not color the frequency response.

The specifications of different rooms of [16] used for our experiments are summarized in Table 1.

**Table 1: Specifications of rooms used for experiments**

| Room | $RT_{60}$ (ms) | Source-microphone setup spacing (meters) | Room dimensions Length × Width × Height (all in meters) | Direct to reverberant ratio (DRR) (dB) |
|---|---|---|---|---|
| X | 0 | 1.5 | 17.04 × 14.53 × 6.5 | ---- |
| A | 320 | 1.5 | 6.6 × 5.7 × 2.3 | 6.09 |
| B | 470 | 1.5 | 4.6 × 4.6 × 2.6 | 5.31 |
| C | 680 | 1.5 | 18.8× 23.5 × 4.6 | 8.82 |
| D | 890 | 1.5 | 8.7 × 8.0 × 4.25 | 6.12 |

### d. Evaluation criteria

We will use two objective evaluation metrics to compare our proposed model performance with other models. These are signal-to-distortion ratio (SDR) [17] and short term objective intelligibility (STOI) [18]. SDR measures the quality and STOI measures the intelligibility of the separated speech as perceived by the listener

### e. Short-time Fourier transform (STFT) parameters

In all experiments, the STFT parameters used are listed in Table 2.

**Table 2: STFT parameters for EM-SONET**

| | |
|---|---|
| Sampling frequency | 16 kHz |
| Window Shape | Hamming |
| STFT frame length | 1024 samples |
| Hop size | 256 samples |
| Velocity of sound | 343 m/s |
| Total EM iterations | 16 |

### f. Different kinds of masks for target retrieval

Different masks can be designed from the ILD soft mask obtained from the pre-trained deep learning neural network and the IPD soft mask obtained from the EM algorithm to retrieve the target. The list is given in Table 3.

In the classical product mask for target retrieval, the ILD and the IPD masks are multiplied for the entire band of speech. In the sub-band mask, a soft mask is prepared by concatenating different portions of ILD and IPD soft masks, according to their strength in different frequency sub-bands of speech. In the weighted sub-band mask, in addition to concatenating different portions of ILD and IPD soft masks; these portions are weighted in order to increase the output performance. As already discussed, MESSL and MESSL-GS algorithms [2] use the product mask and SONET [11] uses the weighted soft-band mask. EM-SONET will use both the product and the sub-band mask in different experiments.

**Table 3: Different masks used for source retrieval in different models used in experiments.**

| Mask for source retrieval | Formation |
|---|---|
| Product mask | Soft ILD mask × Soft IPD mask |
| Sub-band mask | [IPD mask (0 to 1.5 kHz); Product of ILD and IPD masks (1.5 to 4 kHz); ILD mask (4 to 8 kHz)] |
| Weighted sub-band mask | [$β_1$ (IPD masks (0 to 500Hz);$β_2$ (IPD masks (500 to 1.5 kHz);$β_3$ (Product of ILD and IPD masks (1.5 to 4 kHz);$β_4$ (ILD masks (4 to 8 kHz)] |

Where $β_1,……. β_4$ (are the weights calculated empirically for the SONET [11].

### g. Pre-trained network SONET-L

The pre-trained network used in our proposed model is SONET-L. This network is trained on ILD cues of a single source placed in anechoic conditions [11]. In order to reduce the training time and computational cost, it is recommended in [11] to train the U-Net classifier at only few locations in the room. Due to the similarity of interaural cues of the neighboring positions, training at any location is enough for the nearby locations as well. This is termed as the 'generalized solution' in [11]. For example, in case of generalized solution, SONET-L trained at $30^0$ is also used at its neighboring positions ($15^0$ and $45^0$). Likewise, SONET-L trained at $90^0$ is also used for ILD cue classification at its near-neighbors ($60^0$ and $75^0$). Apart from the generalized solution, we will also implement the exclusive solution of SONET-L. The exclusive solution requires a separate pre-trained SONET-L classifier for every target position. The difference between the generalized and exclusive solutions is shown in Figure 4.

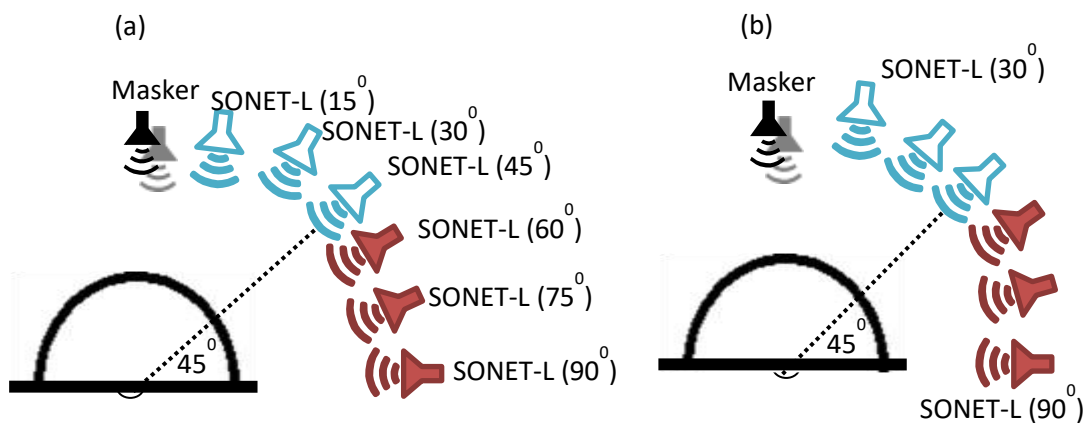

**Figure 4: (a) 'Exclusive' vs. (b) the 'generalized' solution**

The audio mixtures collected at the binaural microphone setup are converted from time domain (one dimension (1D) to TF domain (two dimensions(2D)) signal by using equation (2) and their ILD and IPD spectrograms are extracted using equations 3 and 4. The ILD spectrogram (2D) is treated as a gray scale image, which is given at the input of a pre-trained SONET-L network. SONET-L was trained on the ILD spectrograms of the mixtures collected at the binaural set-up in anechoic conditions. Due to the U-Net input image dimensions (height and width) constraints, the duration of audio clips used to generate the spectrogram for the training phase of SONET-L were so chosen that the dimensions (height and width) of the image must be the multiple of $2^D$, where '$D$' is encoder depth of U-Net [11].

The input image dimensions of the SONET-L network during the training phase were 1024 × 64, but after training it can process any image of dimensions equal to or greater than the images used for training. So when a clean source is convolved with the BRIR of reverberant room, the duration of the source itself and the duration of the audio mixture resulting from the summation of such convolved sources at the microphones would result in images of dimensions greater than the training image dimensions of (1024×64). The exact size depends on the $RT_{60}$ of the room. Greater it is, greater would be the mixture's duration and greater would the resulting ILD spectrogram size (dimensions) in the TF domain. The training parameters of SONET-L are summarized in Table 4.

**Table 4: Training parameters of SONET-L**

| U-Net parameters | Values |
| --- | --- |
| Training samples per class | 31900 |
| Training and testing examples ratio | 0.98/0.02 |
| Input image dimensions (height × width) | 1024 × 64 |
| Optimizer | sgdm (stochastic gradient decent method) |
| L2 regularization | 0.0001 |
| Momentum | 0.95 |
| Initial learning rate | 0.01 |
| Mini batch size | 8 |
| Encoder depth | 2 |
| Stride size of encoder/ decoder convolutional layers | [3,3] |
| Filter size | [3,3] |
| Number of training epochs for SONET-L ($15^0$) | 7 |
| Number of training epochs for SONET-L ($30^0$) | 6 |
| Number of training epochs for SONET-L ($45^0$) | 3 |
| Number of training epochs for SONET-L ($60^0$) | 4 |
| Number of training epochs for SONET-L ($75^0$) | 4 |
| Number of training epochs for SONET-L ($90^0$) | 2 |

### h. Different versions of EM-SONET

We have devised four different versions of EM-SONET on the basis of the mask they are using for target retrieval and whether they are using the generalized or the exclusive solution. The differences among different versions of EM-SONET are mentioned in Table 5.

**Table 5: Different versions of EM-SONET**

| Target masker separations | EM-S(A) | EM-S(B) | EM-S(C) | EM-S(D) |
|---|---|---|---|---|
| **Solution** | Exclusive | Generalized | Exclusive | Generalized |
| **Mask-used** | Product | Product | Sub-band | Sub-band |

### i. Comparison algorithms

We have compared our results with MESSL [2], MESSL-GS (MESSL with garbage source (GS) activated) [2] and SONET [11], in both anechoic and reverberant conditions. SONET model (in this paper) is using the generalized solution and the weighted sub-band mask. The details of these algorithms are mentioned in section II "Related Work".

### j. Solving the EM permutation problem

In BSS evaluation toolkit [17], while calculating SDR, another parameter is generated at the output called 'perm' which depicts the permutation of sources at the model output (if any occurred). If the 'perm' value is {1, 2}, no permutation of sources took place at the EM model's output. However, if its value is {2, 1}, the sources are reversed at the output. We applied the IPD soft masks obtained from the EM algorithm on the mixtures, convert them to time domain and apply BSS evaluation over them to get the value of 'perm' and then combine these soft IPD masks in proper order with their corresponding correct ILD masks according to the 'perm' value. However, note that these BSS evaluations are done just for the purpose of getting the 'perm' value. They do not in any way; represent the proposed model's final output which was calculated after the application of combine ILD-IPD soft masks on the mixtures as described in section "Proposed methodology".

## V. Experimentation Results and Comparison

We have performed five different experiments. In the first experiment, the clustering strength of EM is compared with SONET-P for IPD cues to justify the need of replacing SONET-P with EM in SONET speech separation model proposed in [11]. In second experiment, EM-S(A) is compared with MESSL, MESSL-GS and SONET in anechoic conditions. In the third experiment, we will compare the versions of EM-SONET using different masks. In the fourth experiment, we will compare different versions of EM-SONET on the basis of using the generalized versus the exclusive solution. The insight obtained from the second, third and the fourth experiment is used to select the better version of EM-SONET which would be compared with other algorithms in reverberant conditions in the fifth experiment.

*Case 1:* **Comparison of different approaches for IPD clustering in anechoic conditions**

In this experiment, we will compare the clustering strength of EM with SONET-P to justify the need of replacing SONET-P by EM for clustering the IPD cues, which would finally result in better speech separation. For this purpose, we would turn off the ILD feature of MESSL-GS [2]. The IPD spectrogram are then given as input to SONET-P models (each of them exclusively trained at 15, 30, 60 and 90 degrees separation angles between the target and masker) and MESSL-GS which uses the EM algorithm for clustering the spatial cues. The average quality and intelligibility of target speech in terms of SDR and STOI retrieved from the five speech mixtures in anechoic conditions by these two clustering methods (EM and SONET-P) is shown in Figure 5.

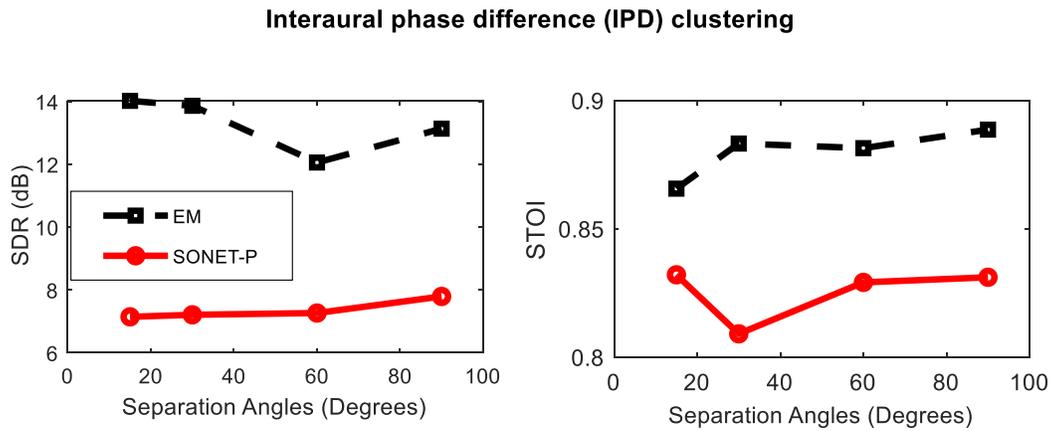

**Figure 5: Performance comparison of different clustering approaches for IPD cues in anechoic conditions ($RT_{60}$ =0ms).**

As shown in Figure 5, the performance of SONET-P at all the separation angles under consideration is lower than the EM algorithm. On average, the EM's retrieved target is better than SONET-P's retrieved target by 6 dB in terms of SDR and by 5% in terms of STOI.

*Case 2:* **Comparison of different algorithms with EM-SONET in anechoic conditions**

In this experiment, we will compare the version (A) of EM-SONET (EM-S(A)) with MESSL, MESSL-GS and SONET. We have used EM-S(A) in this experiment, as it is using the product mask also incorporated in MESSL and MESSL-GS. The results of comparison are given in Figure 6.

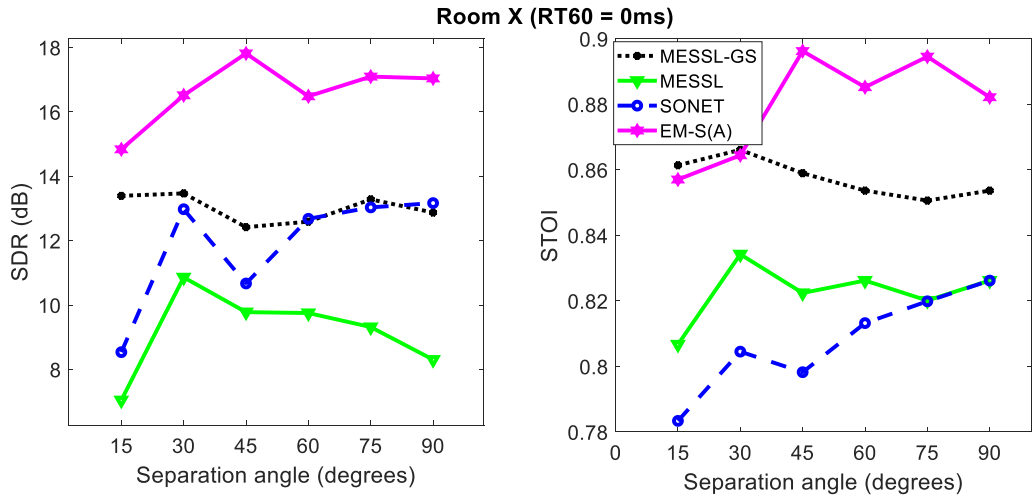

**Figure 6: Performance comparison of EM-SONET with other algorithms in anechoic conditions ($RT_{60}$ =0ms).**

As clear from Figure 6, at all the separation angles, our proposed model EM-SONET (A) performs better than all other algorithms in terms of SDR and STOI except at $15^0$ and $30^0$, where its STOI is lesser than MESSL-GS by 0.6 and 0.3% respectively. When results are averaged over all the angles, our proposed model is better than MESSL by 7.5 dB in terms of SDR and 5.7% in terms of STOI. Similarly, its SDR and STOI are better than MESSL-GS and SONET by 3.6 and 4.8 dB, and 2.2 and 7.2% respectively.

*Case 3:* **Comparison of EM-SONET versions using different masks**

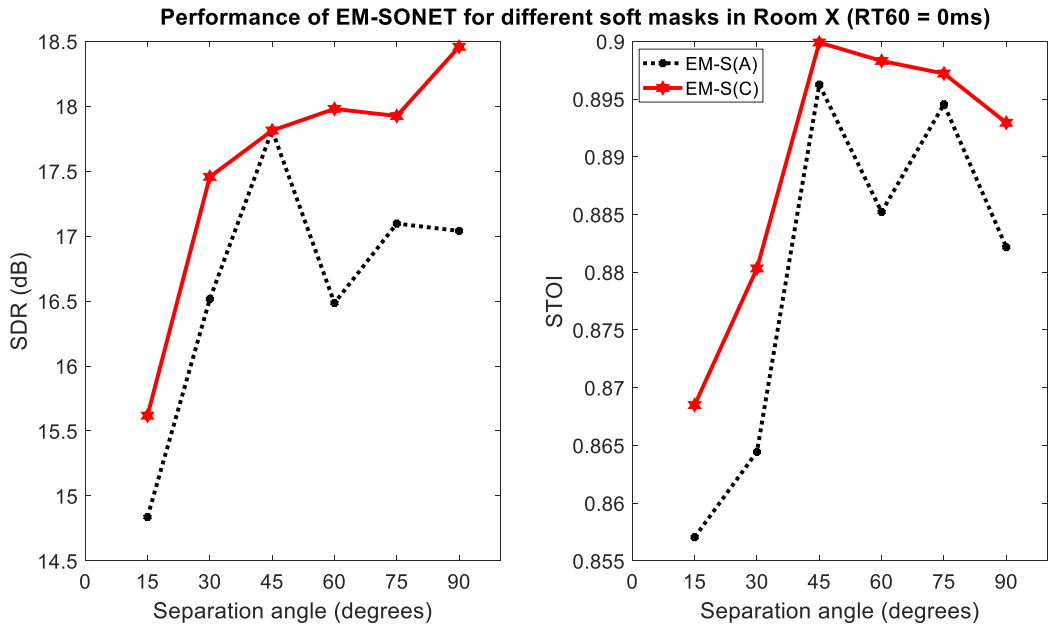

**Figure 7: Performance of EM-SONET at different target-separation angles for the product and the sub-band mask in anechoic conditions ($RT_{60}$ =0ms).**

As stated in [2], the strength of different cues is different in different frequency bands. Utilizing the strength of different cues in different sub-bands, sub-band mask is prepared (See Table 3). In this experiment, the version of EM-SONET using the sub-band mask (EM-S(C)) is compared with the version using the product mask (EM-S(A)). In order to have a fair comparison, both the versions selected for this experiment are using the exclusive solution. As shown in Figure 7, EM-S(C) performs better than EM-S(A). When averaged over all the angles, there is an improvement of 1 dB in SDR and 1% in STOI of EM-S(C) over EM-S(A).

*Case 4:* **Comparison of EM-SONET versions using generalized vs. exclusive solution**

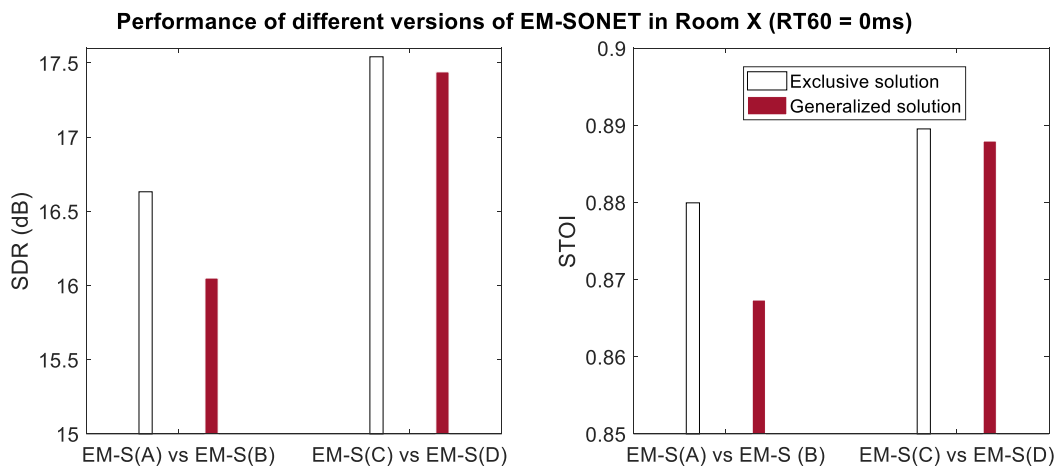

**Figure 8:** Average performance of different versions of EM-SONET for exclusive and generalized solution. The comparison is made between the versions of EM-SONET using similar soft masks as shown in Table 5 in anechoic conditions ($RT_{60}$ =0ms).

As the training phase of neural network is a time consuming process, so it was suggested in [11] to use the 'generalized solution' instead of using the 'exclusive solution'. Versions B and D of EM-SONET (EM-S(B) and EM-S(D)) are using the generalized solution and the versions A and C (EM-S(A) and EM-S(C)) are using the exclusive solution. Figure 8 shows the comparison between different versions of SONET using the generalized and the exclusive solutions. In order to have a fair comparison, the models must be using the same kind of masks (product mask used by EM-S(A) and EM-S(B) and the sub-band masks used by EM-S(C) and EM-S(B)). The results of Figure 8 shows that for the models using the product mask (EM-S(A) vs EM-S(B)), there is an average loss of 0.6 dB in SDR and 1.2% in STOI by using the generalized solution instead of the exclusive solution and negligible loss of 0.1 dB in SDR and 0.1% in STOI for the versions of EM-SONET using the sub-band mask (EM(C) vs. EM(D)). So, a great saving in training time and computational cost can be achieved at a negligible loss of performance by using the generalized solution.

Case 5: **Comparison of different algorithms with EM-SONET in reverberant conditions**

As observed from the three experiments ( $2^{nd}$ , $3^{rd}$ and $4^{th}$ ), EM-S(D) has outperformed other versions of EM-SONET in terms of using better mask (the sub-band mask), bearing lower computational cost and requiring lesser training time. So, on the basis of its previous performance in anechoic conditions, we have selected this version of EM-SONET to compare our proposed model with other algorithms in reverberant conditions. The performance comparisons of EM-S(D) with other algorithms in different reverberant conditions varying from low (room A) to high (room D) are shown in Figures 9 to 12.

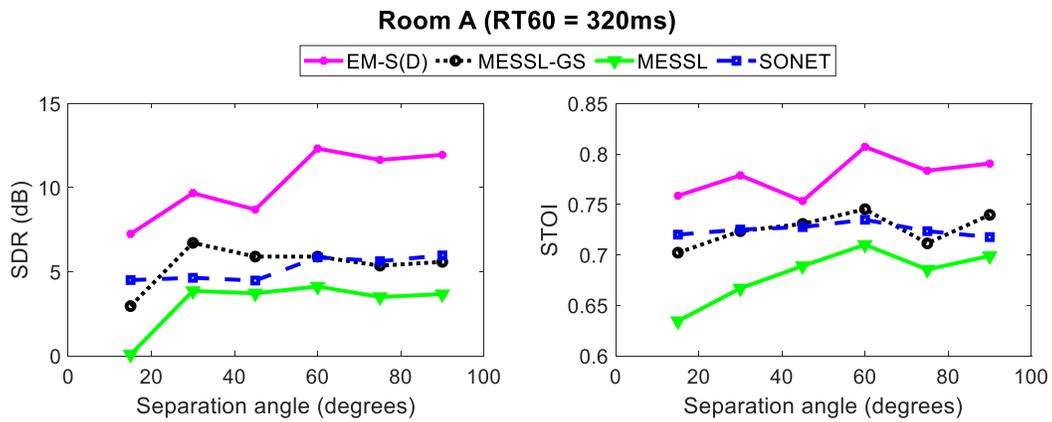

Figure 9: Performance comparison of EM-SONET with other algorithms in room A ($RT_{60}$ =320ms).

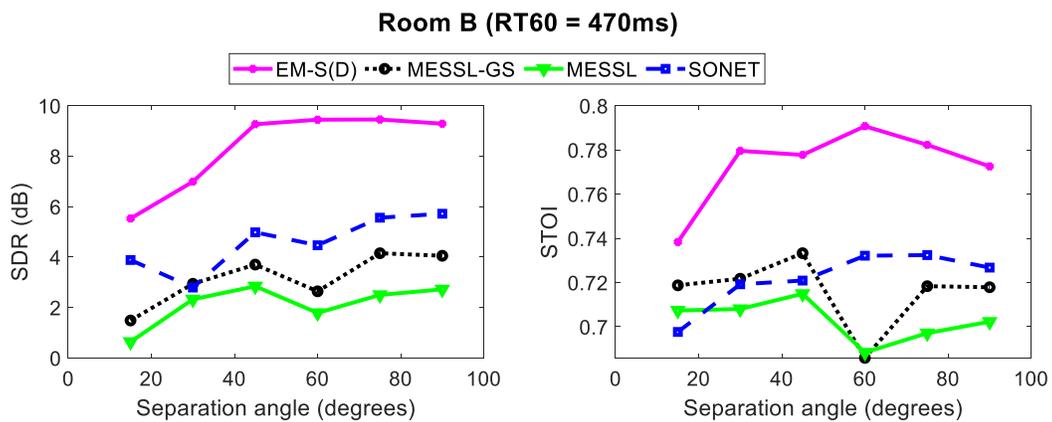

Figure 10: Performance comparison of EM-SONET with other algorithms in room B ($RT_{60}$ =470ms).

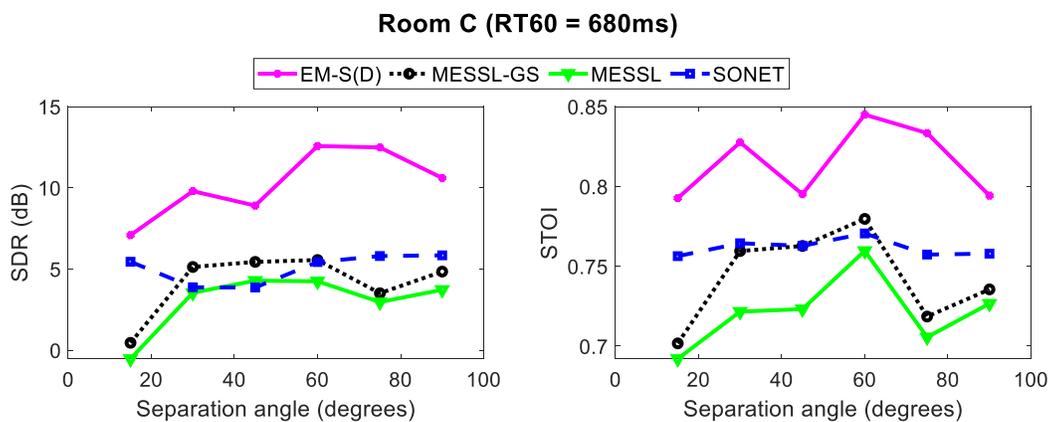

Figure 11: Performance comparison of EM-SONET with other algorithms in room C ($RT_{60}$ =680ms).

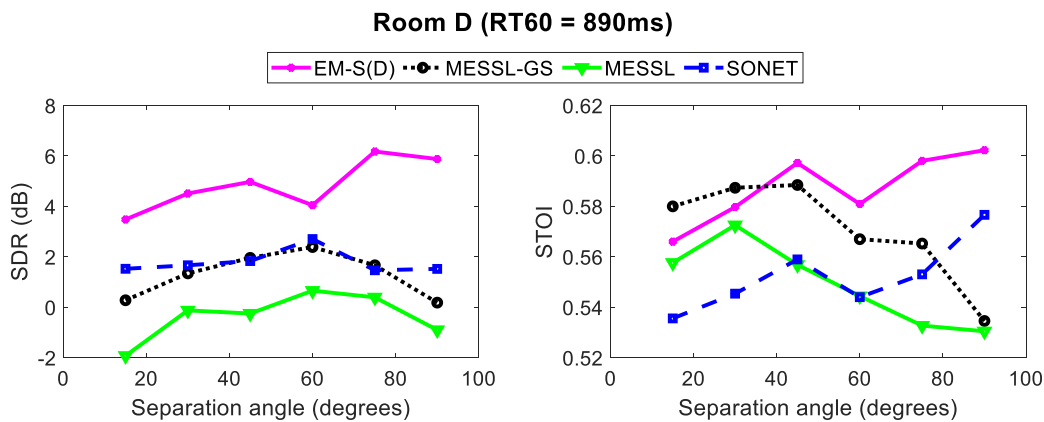

**Figure 12: Performance comparison of EM-SONET with other algorithms in room D ($RT_{60}$ =890ms).**

As clear from these figures, our proposed model EM-S(D) has outperformed other algorithms in all kinds of reverberant conditions. MESSL-GS performs better than MESSL due to the presence of GS, which minimizes the effect of reverberations on system performance by forbidding the outliers to be included and deteriorate the outputs of real sources. Although SONET was designed and trained for anechoic separations, it performs better than MESSL in all the echoic conditions and better than MESSL-GS at few separation angles in terms of SDR, especially the closest one i.e. $15^0$.

In room A, the average SDR of EM-S(D) is higher than MESSL-GS, MESSL and SONET by 4.8, 7.1 and 5.1 dBs and its average STOI is higher by 5.3, 9.8 and 5.4% respectively.

In room B, when averaged over all the angles, the SDR of EM-S(D) is better than MESSL-GS, MESSL and SONET by 5.2, 6.2 and 3.7 dBs and its average STOI is higher by 5.7, 7 and 5.2% respectively.

In room C, the average SDR of EM-S(D) is higher than MESSL-GS, MESSL and SONET by 6.1, 7.2 and 5.2 dBs respectively and its average STOI is better by 7.2, 9.3 and 5.3% respectively.

In room D, the average SDR of EM-S(D) is higher than MESSL-GS, MESSL and SONET by 3.5, 5.2 and 3 dBs respectively and its average STOI is higher by 2, 4 and 3.5% respectively. However, the STOI at smaller separation angles in this room is slightly lesser than MESSL-GS. At $15^0$, the STOI of EM-S is 1.4 % and at $30^0$ it is 0.8% lesser than MESSL-GS. Such small differences are usually perceptually unnoticeable.

Although SONET-L has not encountered the echoes during training, yet the use of deep learning has made possible learning of the direct path ILD cues so perfectly by the network that it has distinguished them easily from the reverberant cues in echoic conditions. SONET-L accompanied by EM has successfully combated the reverberations as shown in Figures 9 to 12.

SONET [11] itself is a great model, as it uses deep learning both for clustering the IPD and ILD cues but the phase wrap issue bring the model down on its knees. Solving this problem by EM algorithm has greatly benefitted the U-Net based speech separation model SONET proposed in [11]. In future, solving this problem by deep learning will definitely revolutionize the speech separation task.

## VI. Conclusion

SONET is a deep neural network based speech separation system which can learn the interaural cues and can outperform the state of the art machine learning speech separation algorithms as shown in [11]. But, its weakness is its failure to cluster the TF units based on IPD cues, due to the phase wrap problem of IPD at higher frequencies, which if removed by the top-down approach of [2] does not work for SONET due to the reduced variance of IPD cues.

This paper was aimed at removing this weakness of SONET by assigning the task of IPD clustering to the machine learning EM algorithm. The results show the proposed hybrid model has achieved better results than the models using either entirely machine learning tools or entirely deep learning tools for interaural speech separation. The use of better mask, reduced computational cost and training time and better performance in different acoustic conditions makes EM-SONET stand out among other deep learning and machine learning algorithms considered in this paper.

**Funding**

This work is supported by Higher Education Commission (HEC), Pakistan, under project no. 6330/KPK/NRPU/R&D/HEC/2016.

**Acknowledgements**

We would like to thank the AI in health care lab, University of Engineering and Technology, Peshawar for sharing their computational facilities. We are also thankful to Michael I. Mandel for sharing his MESSL code.

**References**

[1]. Scott Rickard.: Chapter "The DUET blind source separation algorithm" In book: Blind Speech Separation (pp.217-241), University College, Dublin, Springer 2007.

[2]. Michael I. Mandel, Ron J. Weiss, Daniel P. W. Ellis.: "Model-based expectation-maximization source separation and localization", IEEE Transactions on Audio, Speech and Language Processing, Vol. 18, No. 2, p. 382-394, 2010.

[3]. Kun Han, Deliang Wang.: "An SVM based classification approach to speech separation", Proc. IEEE International Conference on Acoustics, Speech and Signal Processing (ICASSP), 2011, Prague, Czech Republic.

[4]. Paria Dadvar, Masoud Geravanchizadeh.: "Robust binaural speech separation in adverse conditions based on deep neural network with modified spatial features and training target", Speech Communication 108, p. 41–52, 2019.

[5]. Zhong-Qiu Wang, Jonathan Le Roux, John R. Hershey.: "Multi-channel deep clustering: discriminative spectral and spatial embeddings for speaker-independent speech separation", Proc. IEEE International Conference on Acoustics, Speech and Signal Processing (ICASSP), 2018, Alberta, Canada.

[6]. Meet H. Soni, Neil Shah, and Hemant A. Patil.: "Time-frequency masking-based speech enhancement using generative adversarial network", Proc. IEEE International Conference on Acoustics, Speech and Signal Processing (ICASSP), 2018, Alberta, Canada. Paper id- 3043.

[7]. Lukas Drude, Daniel Hasenklever, Reinhold Haeb-Umbach.: "Unsupervised training of a deep clustering model for multichannel blind source separation", Proc. IEEE International Conference on Acoustics, Speech and Signal Processing (ICASSP), 2019, Tokyo, Japan.

[8]. Lukas Drude, Daniel Hasenklever, Reinhold Haeb-Umbach.: "Integration of neural networks and probabilistic spatial models for acoustic blind source separation", IEEE Journal of Selected Topics in Signal Processing Volume: 13, Issue: 4 , 2019.

[9]. Aditya Arie Nugraha, Antoine Liutkus, Member, Emmanuel Vincent.: "Multichannel audio source separation with deep neural networks", IEEE/ACM Transactions on Audio, Speech and Language Processing, Vol. 24, No 9, p. 1652 – 1664, 2016.

[10]. Yi Luo, Zhuo Chen, John R. Hershey, Jonathan Le Roux, Nima Mesgarani.: "Deep clustering and conventional networks for music separation: Stronger together", Proc. IEEE International Conference on Acoustics, Speech and Signal Processing (ICASSP), 2017, https://arxiv.org/abs/1611.06265

[11]. Sania Gul, Muhammad Sheryar Fulaly, Muhammad Salman Khan, Syed Waqar Shah.: "Clustering of spatial cues by semantic segmentation for anechoic source separation", accepted in Applied Acoustics for Volume 171, 1st January 2021 edition. https://doi.org/10.1016/j.apacoust.2020.107566.

[12]. Yang Sun, Waqas Rafique, Jonathon A. Chambers, Syed Mohsen Naqvi.: "Underdetermined source separation using time-frequency masks and an adaptive combined Gaussian-Student's t probabilistic model", Proc. IEEE International Conference on Acoustics, Speech and Signal Processing (ICASSP), 2017, New Orleans, LA, USA.

[13]. Parham Aarabi.: "Self-localizing dynamic microphone arrays", IEEE Transactions on Systems, Man, And Cybernetics—Part C: Applications and Reviews, Vol. 32, No. 4, 2002.

[14]. M. I. Mandel and D. P. W. Ellis.: "EM localization and separation using interaural level and phase cues," Proc. IEEE Workshop Applicat.Signal Process. Audio Acoust., 2007.

[15]. "DAPRA TIMIT acoustic phonetic continuous speech corpus", http://www.ldc.upenn.edu/Catalog/LDC93S1.html

[16]. Christopher Hummersone.: "A psychoacoustic engineering approach to machine sound source separation in reverberant environments", Ph.D. thesis, University of Surrey, 2011.

[17]. Emmanuel Vincent, Remi Gribonval and Cedric Fevotte.: "Performance measurement in blind audio source separation", IEEE Transactions on Audio, Speech And Language Processing, Vol. 14, No. 4, 2006.

[18]. Cees H. Taal, Richard C. Hendriks, Richard Heusdens and Jesper Jensen.: "A short-time objective intelligibility measure for time-frequency weighted noisy speech", Proc. IEEE International Conference on Acoustics, Speech, and Signal Processing (ICASSP), 2010, Dallas, TX, USA.